**Effective manipulation and realization of a colossal nonlinear Hall effect in an electric-field tunable moiré system**

Jinrui Zhong[1,#], Junxi Duan[1,#,*], Shihao Zhang[2,#], Huimin Peng[1], Qi Feng[1], Yuqin Hu[1], Qinsheng Wang[1], Jinhai Mao[3], Jianpeng Liu[2,4,*], Yugui Yao[1,*]

[1]Key Laboratory of Advanced Optoelectronic Quantum Architecture and Measurement (MOE), School of Physics, Beijing Institute of Technology, Beijing 100086, China

[2]School of Physical Science and Technology, ShanghaiTech University, Shanghai 201210, China

[3]School of Physical Sciences and CAS Center for Excellence in Topological Quantum Computation, University of Chinese Academy of Sciences, Beijing 100049, China

[4]ShanghaiTech Laboratory for Topological Physics, ShanghaiTech University, Shanghai 201210, China

\# These authors contributed equally

\* Corresponding authors

E-mails: junxi.duan@bit.edu.cn; liujp@shanghaitech.edu.cn; ygyao@bit.edu.cn

Abstract

The second-order nonlinear Hall effect illuminates a frequency-doubling transverse current emerging in quantum materials with broken inversion symmetry even when time-reversal symmetry is preserved. This nonlinear response originates from both the Berry curvature dipole and the chiral Bloch electron skew scatterings, reflecting various information of the lattice symmetries, band dispersions, and topology of the electron's wavefunctions. Even though many efforts have been put in detecting the nonlinear Hall effect in diverse condensed matter systems, effective manipulation of the two principal mechanisms in a single system has been lacking, and the reported response is relatively weak. Here, we report effective manipulation of the nonlinear Hall effect and realization of a colossal second-order Hall conductivity, $\sim 500~\mu \text{mSV}^{-1}$, orders of magnitudes higher than the reported values, in AB-BA stacked twisted double bilayer graphene. A Berry-curvature-dipole-dominated nonlinear Hall effect, as well as its controllable transition to skew-scattering-dominated response, is identified near the band edge. The colossal response, on the other hand, is detected near the van Hove singularities, mainly determined by the skew scattering of the chiral Bloch electrons. Our findings establish electrically tunable moiré systems promising for nonlinear Hall effect manipulations and applications.

Exotic Hall effect mechanisms have been long-sought in various condensed matter systems as a transport detector of lattice symmetries, electron distributions, and even band topology [1,2]. Recently, a second-order nonlinear Hall effect (NLHE) was proposed that without breaking time-reversal symmetry, a transverse current with double-frequency can be generated by nontrivial Berry curvature distribution in the Brillouin zone, i.e. the Berry curvature dipole (BCD) [3,4]. While a finite value of BCD requires strict symmetry conditions more than inversion symmetry breaking, a second-order nonlinear response has also been proposed to arise from skew scatterings of chiral Bloch electrons [5], in principally allowed in all noncentrosymmetric systems. Noticeably, in some circumstances the skew-scattering mechanism can play a leading role in the nonlinear transport [6,7].

The NLHE not only provides the fundamental information of the topological physics of various Bloch band geometries [8,9], but also inspires the advancement of techniques such as the signal rectification and energy harvest [5,10]. Much effort has been triggered, both theoretically and experimentally [6-8,11-16]. However, although the understanding of the underlying mechanism deepens, effective manipulation of the two principal mechanisms in a single system, which could help excavating the underneath connections between both mechanisms, has been lacking. In addition, the reported second-order nonlinear response in literatures is relatively weak, hampering potential applications of such an intriguing transport effect.

Theories have pointed out that both of the two principal NLHE mechanisms are sensitive to the changes of the band structures and topological properties of the wavefunctions [3,5,11,14]. The twisted graphene systems, therefore, provide a perfect platform for an effective manipulation of the NLHE. With controllable flat bands [17], a series of exotic quantum phenomena such as superconductivity, Chern insulator, and quantum anomalous Hall effect have been detected in them[18]. The study of the nonlinear transport in twisted graphene systems, on the other hand, is still in its early stage [7,9]. Among them, we pick AB-BA stacked twisted double bilayer graphene (TDBG) as an ideal candidate, where the moiré band structure and Fermi energy can be independently tuned [19-24]. As a twisted double bilayer system, the displacement field $D$ not only changes the band dispersions, the charge-neutrality-point (CNP) gap, and the moiré superlattice gaps, but also tunes the topological properties of the moiré flat bands, especially in the AB-BA stacked configuration [25-27]. Meanwhile, highly conductive regimes appear within the flat bands close to the $D$ tunable van Hove singularities (VHS) [22]. In addition, moiré nematic phase has been identified in TDBG [28], together with the fact that the moiré flat bands are susceptible to strain [29], providing multiple symmetry-breaking mechanisms leading to non-zero BCD [30-32].

In this Letter, we report effective manipulation of the two principal NLHE mechanisms and realization of a colossal NLHE response, $\sim 500 \ \mu mSV^{-1}$, orders of magnitudes higher than the reported values, in AB-BA stacked TDBG. The two mechanisms are found to dominate in different regimes of the tuning parameters, i.e. carrier density $n$ and displacement field $D$, with distinct band geometries. Through the independent tuning of $n$ and $D$, we identify that the NLHE undergoes a crossover from the BCD dominant to the skew-scattering dominant near the band edge, achieving an effective manipulation of the two distinct mechanisms in a single system for the first time. Meanwhile, a colossal NLHE response is observed when the chemical potential is tuned to the vicinity of the VHS. In such a situation, the NLHE is determined to be driven by skew scatterings of the chiral Bloch electrons.

The high-quality AB-BA-stacked TDBG device is fabricated with standard "cut-and-stack" technique [Fig. 1(a) and 1(b)] (see Supplemental Material [33]). To achieve AB-BA stacking [Fig. 1(c)], the bottom bilayer graphene was rotated by $60°$ plus a small twisted angle before being picked up. With top and bottom gates, the carrier density $n$ of the graphene channel and the displacement field $D$ perpendicular to it can be independently modulated. Figure 1(d) present the four-probe resistance $R_{xx}$ as a function of $n$ and $D$ from Device 1 measured at 1.7 K. Around $n_s = \pm 4.8 \times 10^{12} \ cm^{-2}$, there are two prominent insulating states induced by the two moiré superlattice gaps, corresponding to 4 electrons/holes per moiré unit cell. From the carrier density $n_s$ required to reach these superlattice gaps, the twisted angle away from $60°$ is determined to be $\theta' = 1.44° \pm 0.05°$ (see Supplemental Material [33]). Below, we will adopt the moiré filling factor $\nu = 4n/n_s$ to denote the position of the chemical potential. The $D$-tunable insulating states at $\nu = 0$ and $\nu = \pm 4$, the cross-like resistivity feature on the hole side, and the insulating states at $\nu = 2$ with 'halo' features surrounding them are similar to the previously reported results [19-23], indicating the high quality of our device. Hall measurement, shown in Fig. 1(e), reveals more features, such as the emerging symmetry-broken states at $\nu =$

3 [22], marked by the $D$-field induced sign change of $R_{xy}$, and the $v$ and $D$ dependence of the VHS, characterized by the diverging Hall density and thus the vanishing $R_{xy}$. Figure 1(f) shows the calculated electronic band structure of the AB-BA stacked TDBG with a twisted angle of $\theta' = 1.44°$. Although the AB-BA stacked TDBG has a band dispersion nearly identical to the one in the AB-AB stacked TDBG, the distribution of the Berry curvature is quite different in the moiré Brillouin zone, resulting in different Chern numbers for the lowest conduction bands [25,26,34]. In addition, when there is strain in TDBG, which is commonly seen in twisted graphene systems [29], the $C_3$ rotation symmetry is broken. Consequently, the distribution of Berry curvature becomes noncentrosymmetric, leading to a nonzero BCD, especially near the band edge [31].

To measure the second-order NLHE, an AC bias current $I^\omega$ with the frequency $\omega$ ($\omega = 17.777$ Hz in all the measurements without specification) was applied between the source and drain of the Hall bar. The measurement setup is sketched in Fig. 1(b). No external magnetic field was applied to preserve the time-reversal symmetry. With lock-in amplifiers, the linear and second-harmonic longitudinal voltages, $V_x^\omega$ and $V_x^{2\omega}$, respectively, and the second-harmonic transverse voltage $V_y^{2\omega}$, were simultaneously recorded. Figure 2(a) shows the $V_y^{2\omega}$ as a function of $v$ and $D$ measured under $I^\omega = 1$ μA at 1.7 K. Referring to Fig. 1(d), $V_y^{2\omega}$ shows peaks and sign change around the insulating states at the CNP gap, the moiré superlattice gap, and the correlation-induced insulating gap at $v = 2$, as shown in Fig. 2(b). To verify that the $V_y^{2\omega}$ is a second-order nonlinear Hall response, we measured $V_y^{2\omega}$ under different excitation current $I^\omega$. Since the resistance changes with varying $v$ and $D$, it is convenient and clear to plot the $V_y^{2\omega}$ against the square of the longitudinal voltage $(V_x^\omega)^2$. Figure 2(c) presents the $V_y^{2\omega} \sim (V_x^\omega)^2$ curves measured at several typical $(v, D)$ points. It is obvious that all the curves are linear. In addition, the direction of $V_y^{2\omega}$ shows no dependence on the direction of the injected current $I^\omega$, presented in Fig. 2(d), and no dependence on the driving frequency (see Supplemental Material Fig. S7 [33]), confirming that it is a second-order NLHE. It should be mentioned that the anomalous Hall effect, despite much weaker than other graphene moiré systems, has been recently reported in both AB-AB and AB-BA stacked TDBG near $v = 3$ and 3.5 [35,36], hinting a possible spontaneous time-reversal-symmetry-broken ground state. However, among graphene moiré systems the spontaneous time-reversal symmetry breaking stems from the orbital magnetism [18,27,34], which is so fragile that it can be easily destroyed by the kinetic-energy enhancement or magneto-electric effect from the driving current [37,38]. In the second-order effect measurements, the applied AC current is typically higher than 100 nA that the spontaneous time-reversal symmetry breaking influence can be ruled out. On the other hand, the AC excitation current was kept less than 1 μA during the whole measurements to avoid any unintended thermoelectric response.

We next explore the microscopic mechanisms of the observed NLHE in AB-BA TDBG. With independently tunable $v$ and $D$, we can study the NLHE in different regimes in TDBG comprised of different contributions. Particularly, $D$ provides an extra knob in TDBG comparing to single-layer graphene and twisted bilayer graphene. Below, we focus on the band edge first, which is believed to host the BCD hotspots [31]. In general, as a second-order effect, the NLHE can be written as $E_y^{2\omega} = \chi_{yxx} E_x E_x / \sigma$, where $E_y^{2\omega} = V_y^{2\omega}/W$ is the second-order transverse electric field driven by the longitudinal electric field $E_x = V_x/L$, $\chi_{yxx}$ the second-order Hall conductivity, $L$ and $W$ the length and width of the Hall bar, respectively. Figure 3(a) shows $E_y^{2\omega}/(E_x^{\omega})^2$ and $\sigma$ as a function of $D$ at $v = -0.188$. As $D$ increases, $\sigma$ keeps decreasing. $E_y^{2\omega}/(E_x^{\omega})^2$, however, shows quite different and complicate behavior. Since it becomes zero when $D > 0.3$ V/nm, we will concentrate in the $D \leq 0.3$ V/nm regime. According to theory, when both BCD-induced intrinsic and impurity contributions are considered, the scaling law between $\chi_{yxx}$ and $\sigma$ is written as

$$\frac{\chi_{yxx}}{\sigma} = \frac{E_y^{2\omega}}{(E_x^{\omega})^2} = [C_1\sigma_0^{-1} + (C_2 - C_3 + C_4)\sigma_0^{-2}]\sigma^2 + (C_3 - 2C_4)\sigma_0^{-1}\sigma^1 + C_4 \quad (1)$$

where $C_{1,2,3,4}$ are the scaling parameters contributed by different sources (see Supplemental Material [33]), $\sigma_0$ the zero-temperature conductivity [11,39]. In the present case, considering the limited thermal excitations at 1.7 K, $\sigma$ can be replaced by $\sigma_0$. As a result, Eq. 1 reduces to $E_y^{2\omega}/(E_x^{\omega})^2 = C_1\sigma_0 + C_2$. We therefore plot $E_y^{2\omega}/(E_x^{\omega})^2$ against $\sigma$ at several $v$ in Fig. 3(b). All the curves show quite similar $\sigma$ dependence. From higher $\sigma$ to lower $\sigma$, they can be divided into three sections. In section I (high $\sigma$), $E_y^{2\omega}/(E_x^{\omega})^2$ is generally independent of $\sigma$, indicating that the NLHE is mainly determined by $C_2$. In section II (medium $\sigma$), $E_y^{2\omega}/(E_x^{\omega})^2$ increases linearly with decreasing $\sigma$. This linear dependence shows that $C_1$ starts to play a role, but the opposite sign between $C_1$ and $E_y^{2\omega}/(E_x^{\omega})^2$ means that the NLHE in this section is still dominated by $C_2$. The values of $C_1$ and $C_2$ can be extracted from the linear fitting. In section III (low $\sigma$), $E_y^{2\omega}/(E_x^\omega)^2$ rapidly drops to zero with further decrease of $\sigma$. This can be explained by the impurity band induced by disorders in which the charge carriers are localized and the variable-range hopping mechanism dominates the conduction [40-42].

From the theory [11], $C_1$ represents the non-Gaussian skew scattering off impurities (see Supplemental Material [33]). $C_2$, on the other hand, is comprised of three contributions, including the intrinsic, the side-jump, and the Gaussian skew scattering off impurities. To get a deeper insight of the mechanisms, we plot the $C_1$ and $C_2$ extracted from the curves as a function of $v$. Figure 3(c) shows the extracted $C_1$ and $C_2$. In section I, $C_1 \sim 0$ such that the non-Gaussian skew scattering off impurities is negligible. The NLHE is determined by $C_2$. As the hole density decreases, the magnitude of $C_2$ gradually increases, consistent with the fact that the band edge hosts the BCD hotspots, implying that $C_2$ is mainly contributed by the intrinsic mechanism. If we neglect the other contributions to $C_2$, we can obtain an estimation of the magnitude of BCD from $C_2$ (see Supplemental Material [33]), plotted in Fig. 3(d). In section II, $C_1$ is nonzero, indicating that disorders play an important role in this section. For $C_2$, under the same hole density, it is larger in section II than in section I. In addition, in section II the magnitudes of both $C_1$ and $C_2$ increase with the increase of the hole density. Both features, together with the opposite $v$ dependence of $C_2$ in section II to the one in section I, suggest that the Gaussian skew-scattering contribution in $C_2$ is no longer negligible in section II but can even be dominant. Unlike the intrinsic mechanism having its maximum around the band edge, the skew-scattering contribution increases as the carrier density increases and becomes the strongest contribution at higher carrier density [11]. The nonlinear Hall signal measured from another $\theta' \approx 1.05°$ TDBG sample shows almost identical trend (see Supplemental Material Fig. S2 [33]).

To collaborate with the scaling results, we calculate the band structure of the TDBG under several different values of interlayer electrostatic potential drop $\Delta$ and the corresponding BCD by introducing a 0.1% uniaxial strain into the system (see Supplemental Material [33]). Here, we should mention that, although moiré nematicity breaks the $C_3$ symmetry, it is prominent only in a doping range away from CNP [28]. Figure 3(e) shows the dependence of the $x$ component of the BCD ($\Lambda_x$) on the Fermi energy with $\Delta = 10$ meV, corresponding to $D \approx$

0.12 V/nm, where the interlayer distance $d = 0.335$ nm and the effective dielectric constant $\varepsilon = 4$ were adopted. It is obvious that $\Lambda_x$ peaks around the CNP where the edges of the conduction and valence bands overlap with each other. As the Fermi level moves to the hole side, $\Lambda_x$ drops rapidly and then reverse its sign. It agrees with the decrease of experimentally evaluated BCD with increasing hole density in section I and the sign reversal of $E_y^{2\omega}/(E_x^{\omega})^2$ at higher hole density [Fig. 2(a)]. More importantly, the calculated BCD perfectly matches the value estimated from the intercept ($C_2$) in section I. All these features strongly hint that the NLHE in section I is dominated by the intrinsic mechanism. We also calculate the dependence of $\Lambda_x$ on the Fermi energy with $\Delta = 15$ meV, plotted in Fig. 3(f), corresponding to $D \approx 0.18$ V/nm, thus in section II. Similar to Fig. 3(e), $\Lambda_x$ decreases with increasing hole density. However, the experimentally extracted $C_2$ increases with increasing hole density in section II. In addition, at the same hole density, $\Lambda_x$ under higher $D$ is smaller than the one under lower $D$. This is opposite to the fact that the experimentally measured $C_2$ is larger in section II than the one in section I. Both contradictions indicate that $C_2$ in section II is no longer dominated by the intrinsic mechanism, rather the skew-scattering mechanisms play a leading role. Therefore, the theoretical calculations strongly support the experimental results that we have effectively manipulated the NLHE to undergo a transition between the two principal NLHE mechanisms.

Finally, we will discuss the colossal $\chi_{yxx}$ observed in TDBG. As shown by the theory [11,39], $\chi_{yxx}$ strongly depends on $\sigma$, especially for the disorder-related mechanisms. Figure 4(a) plots the $v$ and $D$ dependence of $\chi_{yxx}$ measured at 1.7 K under $I^\omega = 1$ μA. There are several regimes where $\chi_{yxx}$ is larger than 100 μmSV[-1] which is an order of magnitude higher than the record values in reported studies (see Supplemental Material Table S1 [33]). To understand the origin of this colossal $\chi_{yxx}$, we measured the NLHE under different temperatures. Several $(v, D)$ points in different regimes were selected, and we will focus on the results from one of them below (see Supplemental Material [33]). Figure 4(b) plots $E_y^{2\omega}$ as a function of $(E_x^{\omega})^2$ at $v = -2.5$ and $D = +0.56$ V/nm under different temperatures. It is obvious that all the curves can be well captured by a linear fitting. We then extract the temperature dependence of $\chi_{yxx}$ and $\sigma$, shown in Fig. 4(c). Both decrease with increasing temperature. A careful analysis in Fig. 4(d) shows that $\chi_{yxx}$ scales linearly against $\sigma^3$ with a zero intercept. This holds true for the other points (see Supplemental Material Fig. S3 [33]). According to the scaling law (Eq. 1), this scaling behavior indicates that the colossal NLHE is dominated by skew scattering mechanism (see Supplemental Material [33]). The scattering process arises from the theoretically proposed skew scattering of the chiral Bloch electrons [5]. The Bloch wave functions in the valley $K$ and $K'$ have opposite momentum-dependent chirality and Berry curvature with time-reversal symmetry [43]. In the condition of broken inversion symmetry, an asymmetric scatterings process of these chiral Bloch electrons could emerge, leading to a finite nonlinear transverse transport. This skew-scattering response has a $\sigma^3$ dependence, and thus the ultra-high conductivity accounts for the observed colossal NLHE.

It is interesting to note that the mechanism of the colossal NLHE here is quite similar to the one reported in the aligned h-BN/graphene superlattice devices [6], but is totally different from our previous results observed in twisted bilayer graphene (TBG) [7]. In the TBG devices, phonon skew scattering plays an important role in the NLHE, with its contribution comparable to the one from impurity skew scattering. However, as shown by the scaling results presented above, phonon skew scattering is negligible in TDBG. A possible reason for this difference is that the peak response of the NLHE is observed close to the superlattice gaps in TBG but away from gaps in TDBG and h-BN/graphene devices. Understanding this difference is important for physical research as well as potential room-temperature applications of the NLHE, thus further experimental and theoretical works are called.

To summarize, through dual-gate technique, we have succeeded in manipulating the two principal mechanisms, BCD and skew scattering, to dominate the NLHE respectively in distinct $v$ and $D$ regimes. Near the band edges, the NLHE performs a transition from the BCD dominant to the skew-scattering dominant. We have also detected a record-high second-order Hall conductivity $\chi_{yxx}$ over $500\ \mu\text{mSV}^{-1}$, orders larger than previously reported values. The colossal response exists in certain regimes near the VHS, dominated by skew scattering of chiral Bloch electrons induced by impurities. Our results establish electrically tunable moiré systems for NLHE mechanism modulation, pushing forward the subsequent research efforts on this exotic Hall effect mechanism and its potential applications.

**References**

[1] F. D. M. Haldane, Berry curvature on the fermi surface: anomalous Hall effect as a topological fermi-liquid property, Phys. Rev. Lett. **93**, 206602 (2004).

[2] N. Nagaosa, J. Sinova, S. Onoda, A. H. MacDonald, and N. P. Ong, Anomalous Hall effect, Rev. Mod. Phys. **82**, 1539 (2010).

[3] I. Sodemann and L. Fu, Quantum Nonlinear Hall Effect Induced by Berry Curvature Dipole in Time-Reversal Invariant Materials, Phys. Rev. Lett. **115**, 216806 (2015).

[4] Z. Z. Du, H.-Z. Lu, and X. C. Xie, Nonlinear Hall effects, Nat. Rev. Phys. **3**, 744 (2021).

[5] H. Isobe, S.-Y. Xu, and L. Fu, High-frequency rectification via chiral Bloch electrons, Sci. Adv. **6**, eaay2497 (2020).

[6] P. He, G. K. W. Koon, H. Isobe, J. Y. Tan, J. Hu, A. H. C. Neto, L. Fu, and H. Yang, Graphene moiré superlattices with giant quantum nonlinearity of chiral Bloch electrons, Nat. Nanotechnol. **17**, 378 (2022).

[7] J. Duan, Y. Jian, Y. Gao, H. Peng, J. Zhong, Q. Feng, J. Mao, and Y. Yao, Giant Second-Order Nonlinear Hall Effect in Twisted Bilayer Graphene, Phys. Rev. Lett. **129**, 186801 (2022).

[8] Q. Ma, S. Y. Xu, H. Shen, D. MacNeill, V. Fatemi, T. R. Chang, A. M. Mier Valdivia, S. Wu, Z. Du, C. H. Hsu, S. Fang, Q. D. Gibson, K. Watanabe, T. Taniguchi, R. J. Cava, E. Kaxiras, H. Z. Lu, H. Lin, L. Fu, N. Gedik, and P. Jarillo-Herrero, Observation of the nonlinear Hall effect under time-reversal-symmetric conditions, Nature **565**, 337 (2019).

[9] S. Sinha, P. C. Adak, A. Chakraborty, K. Das, K. Debnath, L. D. V. Sangani, K. Watanabe, T. Taniguchi, U. V. Waghmare, A. Agarwal, and M. M. Deshmukh, Berry curvature dipole senses topological transition in a moiré superlattice, Nat. Phys. **18**, 765 (2022).

[10] Y. Zhang and L. Fu, Terahertz detection based on nonlinear Hall effect without magnetic field, Proc. Natl. Acad. Sci. U.S.A **118** (2021).

[11] Z. Z. Du, C. M. Wang, S. Li, H. Z. Lu, and X. C. Xie, Disorder-induced nonlinear Hall effect with time-reversal symmetry, Nat. Commun. **10**, 3047 (2019).

[12] K. Kang, T. Li, E. Sohn, J. Shan, and K. F. Mak, Nonlinear anomalous Hall effect in few-layer $WTe_2$, Nat. Mater. **18**, 324 (2019).

[13] P. He, H. Isobe, D. Zhu, C. H. Hsu, L. Fu, and H. Yang, Quantum frequency doubling in the topological insulator $Bi_2Se_3$, Nat. Commun. **12**, 698 (2021).

[14] Z. Z. Du, C. M. Wang, H. P. Sun, H. Z. Lu, and X. C. Xie, Quantum theory of the nonlinear Hall effect, Nat. Commun. **12**, 5038 (2021).

[15] A. Tiwari, F. Chen, S. Zhong, E. Drueke, J. Koo, A. Kaczmarek, C. Xiao, J. Gao, X. Luo, Q. Niu, Y. Sun, B. Yan, L. Zhao, and A. W. Tsen, Giant c-axis nonlinear anomalous Hall effect in Td-$MoTe_2$ and $WTe_2$, Nat. Commun. **12**, 2049 (2021).

[16] M. Huang, Z. Wu, J. Hu, X. Cai, E. Li, L. An, X. Feng, Z. Ye, N. Lin, K. T. Law, and N. Wang, Giant nonlinear Hall effect in twisted bilayer $WSe_2$, National Science Review (2022).

[17] R. Bistritzer and A. H. MacDonald, Moire bands in twisted double-layer graphene, Proc. Natl. Acad. Sci. U.S.A **108**, 12233 (2011).

[18] E. Y. Andrei and A. H. MacDonald, Graphene bilayers with a twist, Nat. Mater. **19**, 1265 (2020).

[19] Y. Cao, D. Rodan-Legrain, O. Rubies-Bigorda, J. M. Park, K. Watanabe, T. Taniguchi, and P. Jarillo-Herrero, Tunable correlated states and spin-polarized phases in twisted bilayer-bilayer

graphene, Nature **583**, 215 (2020).

[20] X. Liu, Z. Hao, E. Khalaf, J. Y. Lee, Y. Ronen, H. Yoo, D. Haei Najafabadi, K. Watanabe, T. Taniguchi, A. Vishwanath, and P. Kim, Tunable spin-polarized correlated states in twisted double bilayer graphene, Nature **583**, 221 (2020).

[21] C. Shen, Y. Chu, Q. Wu, N. Li, S. Wang, Y. Zhao, J. Tang, J. Liu, J. Tian, K. Watanabe, T. Taniguchi, R. Yang, Z. Y. Meng, D. Shi, O. V. Yazyev, and G. Zhang, Correlated states in twisted double bilayer graphene, Nat. Phys. **16**, 520 (2020).

[22] M. He, Y. Li, J. Cai, Y. Liu, K. Watanabe, T. Taniguchi, X. Xu, and M. Yankowitz, Symmetry breaking in twisted double bilayer graphene, Nat. Phys. **17**, 26 (2020).

[23] L. Liu, S. Zhang, Y. Chu, C. Shen, Y. Huang, Y. Yuan, J. Tian, J. Tang, Y. Ji, R. Yang, K. Watanabe, T. Taniguchi, D. Shi, J. Liu, W. Yang, and G. Zhang, Isospin competitions and valley polarized correlated insulators in twisted double bilayer graphene, Nat. Commun. **13**, 3292 (2022).

[24] Q. Li, B. Cheng, M. Chen, B. Xie, Y. Xie, P. Wang, F. Chen, Z. Liu, K. Watanabe, T. Taniguchi, S. J. Liang, D. Wang, C. Wang, Q. H. Wang, J. Liu, and F. Miao, Tunable quantum criticalities in an isospin extended Hubbard model simulator, Nature **609**, 479 (2022).

[25] M. Koshino, Band structure and topological properties of twisted double bilayer graphene, Phys. Rev. B **99**, 235406 (2019).

[26] N. R. Chebrolu, B. L. Chittari, and J. Jung, Flat bands in twisted double bilayer graphene, Phys. Rev. B **99**, 235417 (2019).

[27] J. Liu, Z. Ma, J. Gao, and X. Dai, Quantum Valley Hall Effect, Orbital Magnetism, and Anomalous Hall Effect in Twisted Multilayer Graphene Systems, Phys. Rev. X **9**, 031021 (2019).

[28] C. Rubio-Verdú, S. Turkel, Y. Song, L. Klebl, R. Samajdar, M. S. Scheurer, J. W. F. Venderbos, K. Watanabe, T. Taniguchi, H. Ochoa, L. Xian, D. M. Kennes, R. M. Fernandes, Á. Rubio, and A. N. Pasupathy, Moiré nematic phase in twisted double bilayer graphene, Nat. Phys. **18**, 196 (2022).

[29] N. P. Kazmierczak, M. Van Winkle, C. Ophus, K. C. Bustillo, S. Carr, H. G. Brown, J. Ciston, T. Taniguchi, K. Watanabe, and D. K. Bediako, Strain fields in twisted bilayer graphene, Nat. Mater. **20**, 956 (2021).

[30] P. A. Pantaleón, T. Low, and F. Guinea, Tunable large Berry dipole in strained twisted bilayer graphene, Phys. Rev. B **103**, 205403 (2021).

[31] C.-P. Zhang, J. Xiao, B. T. Zhou, J.-X. Hu, Y.-M. Xie, B. Yan, and K. T. Law, Giant nonlinear Hall effect in strained twisted bilayer graphene, Phys. Rev. B **106**, L041111 (2022).

[32] R. Battilomo, N. Scopigno, and C. Ortix, Berry Curvature Dipole in Strained Graphene: A Fermi Surface Warping Effect, Phys. Rev. Lett. **123**, 196403 (2019).

[33] See Supplemental Material for details of device fabrication, determine the twisted angle, transport measurements, second-order nonlinear conductivity and the scaling law, estimate the Berry curvature dipole, theoretical calculation of the band structure and Berry curvature dipole, comparison of second-order nonlinear conductivities $\chi_{yxx}$ from different works, data from TDBG Device 2, extra data from Device 1, effect of driving frequency.

[34] J. Liu and X. Dai, Orbital magnetic states in moiré graphene systems, Nat. Rev. Phys. **3**, 367 (2021).

[35] M. Kuiri, C. Coleman, Z. Gao, A. Vishnuradhan, K. Watanabe, T. Taniguchi, J. Zhu, A. H.

MacDonald, and J. Folk, Spontaneous time-reversal symmetry breaking in twisted double bilayer graphene, Nat. Commun. **13**, 6468 (2022).

[36] M. He, J. Cai, Y.-H. Zhang, Y. Liu, Y. Li, T. Taniguchi, K. Watanabe, D. H. Cobden, M. Yankowitz, and X. Xu, Chirality-dependent topological states in twisted double bilayer graphene, arXiv:2109.08255.

[37] M. Serlin, C. L. Tschirhart, H. Polshyn, Y. Zhang, J. Zhu, K. Watanabe, T. Taniguchi, L. Balents, and A. F. Young, Intrinsic quantized anomalous Hall effect in a moiré heterostructure, Science **367**, 900 (2019).

[38] A. L. Sharpe, E. J. Fox, A. W. Barnard, J. Finney, K. Watanabe, T. Taniguchi, M. A. Kastner, and D. Goldhaber-Gordon, Emergent ferromagnetism near three-quarters filling in twisted bilayer graphene, Science **365**, 605 (2019).

[39] C. Xiao, H. Zhou, and Q. Niu, Scaling parameters in anomalous and nonlinear Hall effects depend on temperature, Phys. Rev. B **100**, 161403(R) (2019).

[40] J. B. Oostinga, H. B. Heersche, X. Liu, A. F. Morpurgo, and L. M. K. Vandersypen, Gate-induced insulating state in bilayer graphene devices, Nat. Mater. **7**, 151 (2008).

[41] T. Taychatanapat and P. Jarillo-Herrero, Electronic transport in dual-gated bilayer graphene at large displacement fields, Phys. Rev. Lett. **105**, 166601 (2010).

[42] E. McCann and M. Koshino, The electronic properties of bilayer graphene, Reports on Progress in Physics **76**, 056503 (2013).

[43] D. Xiao, W. Yao, and Q. Niu, Valley-contrasting physics in graphene: magnetic moment and topological transport, Phys. Rev. Lett. **99**, 236809 (2007).

**Acknowledgement**

The research was supported by the National Key R&D Program of China (Grants No. 2020YFA0308800, No. 2019YFA0308402, No. 2020YFA0309601), the National Natural Science Foundation of China (Grants No. 12234003, No. 12061131002, No. 12174257, No. 61804008), the Beijing Natural Science Foundation (Grant No. Z190006), the Strategic Priority Research Program of Chinese Academy of Sciences (Grant No. XDB30000000). The fabrication was supported by Micro-nano center of Beijing Institute of Technology.

**Figures and captions**

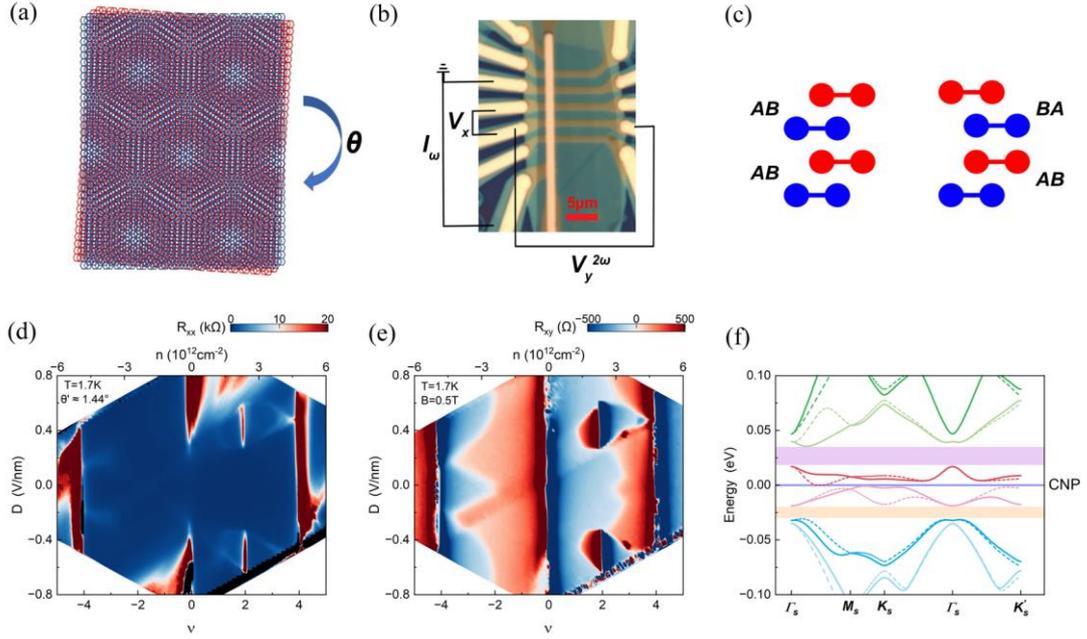

FIG. 1. (a) A schematic of moiré superlattices. Two Bernal-stacked bilayer graphene are twisted by a small angle $\theta$, creating a long-range moiré pattern. (b) A microscopic optical image of the AB-BA stacked $60°+1.44°$ TDBG Hall bar device with Ti/Cr/Au top gate and graphite bottom gate. An AC current $I^\omega$ is applied. The longitudinal and transverse voltage drops are measured simultaneously. The scale bar is 5 μm. (c) The two stacking types of TDBG. An AB sheet denotes a single bilayer graphene. (d) Longitudinal resistance $R_{xx}$ as a function of filling $\nu$ and displacement field $D$ measured at temperature T=1.7 K with zero magnetic field. (e) Linear Hall resistance $R_{xy}$ as a function of filling $\nu$ and displacement field $D$ measured at temperature T=1.7 K with a magnetic field B=0.5 T. (f) Bloch band dispersion of AB-BA stacked $60°+1.44°$ TDBG calculated by continuum model with interlayer potential $\Delta=$ 5 meV. The solid and dashed lines are bands for $k$ valley and $k'$ valley, respectively. The single-particle bandgaps are highlighted by purple and orange bars. The blue bar stands for the charge neutral point (CNP).

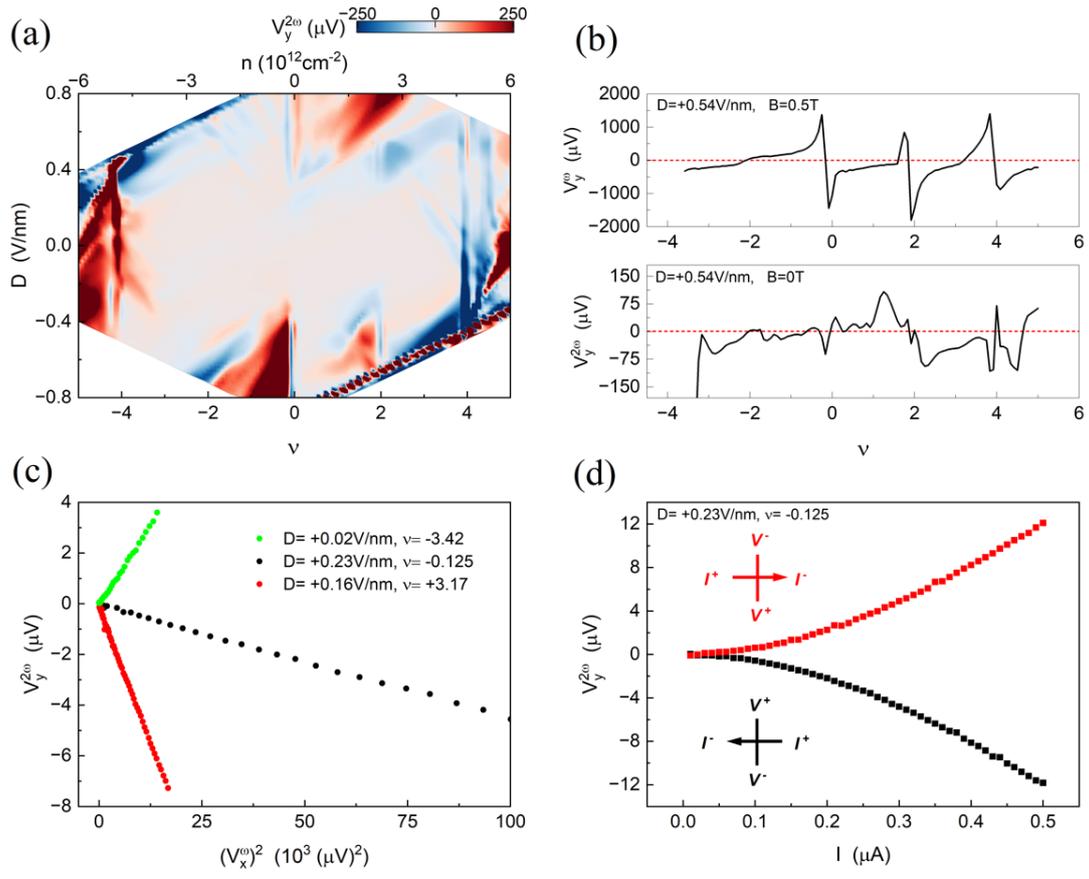

FIG. 2. (a) Second-order nonlinear Hall voltage $V_y^{2\omega}$ as a function of filling factor $\nu$ and displacement field $D$ measured by applying an AC current $I^\omega = 1$ µA with a frequency $\omega = 17.777$ Hz at temperature T=1.7 K. (b) The lower panel, $V_y^{2\omega}$ as a function of $\nu$ at a fixed $D = +0.54$ V/$nm$ extracted from (a); the upper panel, linear Hall voltage $V_y^\omega$ at the same fixed $D$, by multiply $I^\omega = 1$ µA with $R_{xy}$ extracted from Fig. 1(e). The upper panel is for comparing the sign change of carrier density. (c) $V_y^{2\omega}$ versus the square of longitudinal voltage $(V_x^\omega)^2$ at several chosen $\nu$ and $D$ measured by changing $I^\omega$ at 1.7 K. (d) $V_y^{2\omega}$ at $\nu = -0.125$ and $D = +0.23$ V/$nm$ versus current $I$ with two types of current direction at 1.7 K. The black square and red circular curves have different driving current direction and Hall measurement geometry as shown by the crisscross symbols.

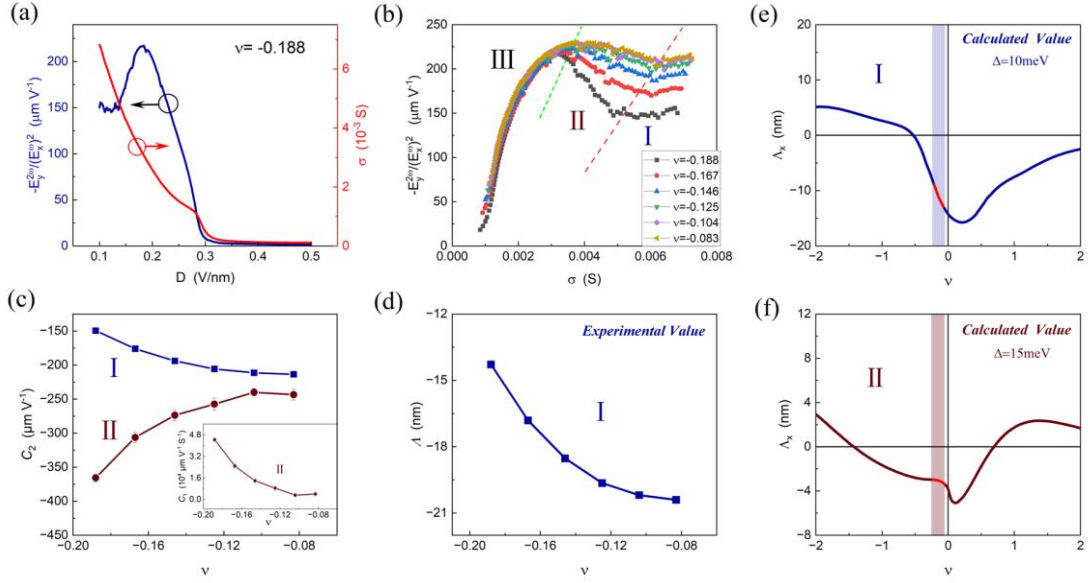

FIG. 3. (a) The nonlinear coefficient $-E_y^{2\omega}/(E_x^\omega)^2$ (blue curve, left axis) and conductivity $\sigma$ (red curve, right axis) change with $D$ at fixed filling $\nu = -0.188$. The applied current is $I^\omega = 1$ μA, and the temperature is T=1.7 K. $E_y^{2\omega}$ and $E_x^\omega$ are the second-order transverse and linear longitudinal electric field, respectively. (b) $-E_y^{2\omega}/(E_x^\omega)^2$ versus $\sigma$ by changing $D$ at various fixed $\nu$ near CNP. These curves are divided into three sections marked as I, II and III according to their trend. The green and red dashed lines are plotted to roughly distinguish the section borders. It should be noticed that, we use the minus value for $E_y^{2\omega}/(E_x^\omega)^2$ in (a) and (b) to display a clearer view of the trend. (c) The extracted scaling coefficient $C_2$ as a function of filling $\nu$ in section I (blue square line) and II (claret circle line). The inset shows the extracted $C_1$ in section II. (d) Estimated BCD value $\Lambda$ from $C_2$ for section I. The BCD estimating process is in Supplemental Material. (e) and (f) Theoretically calculated BCD $\Lambda_x$ (see Supplemental Material) as a function of $\nu$ with the interlayer potential $\Delta = 10$ meV (e, blue line, section I) and 15meV (f, claret line, section II), corresponding to $D \approx 0.12$ V/nm and 0.18 V/nm, respectively. The shadow area in (e) and (f) denotes the experimental filling range in (d).

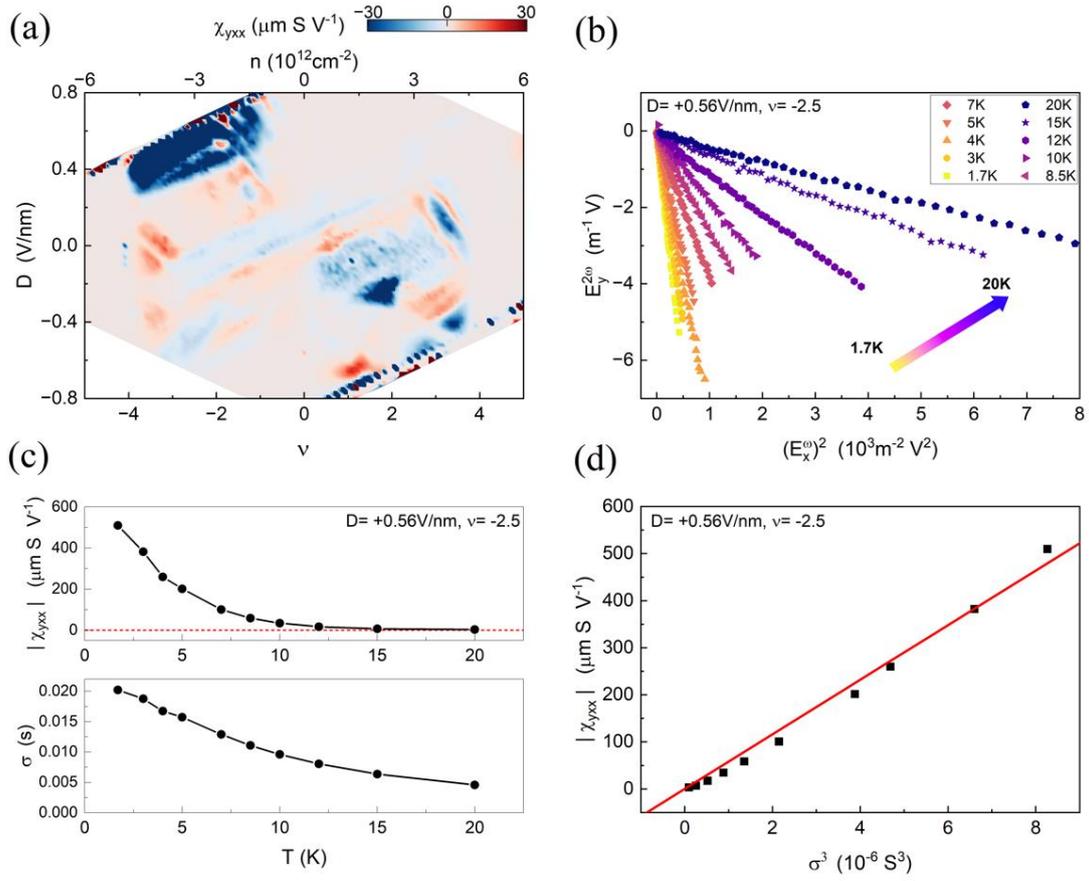

FIG. 4. (a) The second-order nonlinear conductivity $\chi_{yxx}$ as a function of filling $\nu$ and displacement field $D$ measured by applying an AC current $I^\omega = 1$ μA at 1.7 K. (b) The second-order transverse electric field $E_y^{2\omega}$ versus the square of linear longitudinal electric field $(E_x^\omega)^2$ of $D = +0.56$ V/$nm$ and $\nu = -2.5$ at various temperatures. (c) Temperature dependence of the second-order nonlinear conductivity $|\chi_{yxx}|$ (the upper panel) and linear conductivity $\sigma$ (the lower panel). The red dashed line is for guiding to zero. (d) The scaling relation $|\chi_{yxx}| \sim \sigma^3$ between $|\chi_{yxx}|$ and $\sigma$ extracted from (c). The red line is the zero-intercept linear fitting line. The error bar of each data point in (c) and (d) is smaller than the size of the scatters.